\documentstyle[aps,twocolumn,graphicx,ifthen]{revtex}

\newcommand{\myscalebox}[1]{\scalebox{0.45}[0.5]{#1}}
\newcommand{\myscaleboxb}[1]{\scalebox{0.6}[0.6]{#1}}
\newcommand{\myscaleboxc}[1]{\scalebox{0.5}[0.5]{#1}}
\newcommand{\mysection}[1]{\section*{ #1}}
\begin{document}
\draft

\title{Ordered phase in the two-dimensional randomly coupled ferromagnet}

\author{A.K. Hartmann 
\thanks{hartmann@theorie.physik.uni-goettingen.de
}
Institut f\"ur theoretische Physik, Bunsenstr. 9, 37073 G\"ottingen, Germany}

\author{I.A. Campbell
\thanks{campbell@LDV.univ-montp2.fr}
Laboratoire des Verres, Universit\'e Montpellier II, Place
Eug\`ene Bataillon,  34095 Montpellier Cedex 05, France, and Physique des
Solides, Universit\'e Paris Sud, 91405 Orsay, France}

\date{\today}
\maketitle
\begin{abstract}

True ground states are evaluated for a $2d$ Ising model with random near
neighbor interactions and ferromagnetic second neighbor interactions (the
Randomly Coupled Ferromagnet). The  spin glass stiffness exponent is
positive when the absolute value of the random interaction is weaker than
the ferromagnetic interaction. This result demonstrates that in this
parameter domain the spin glass like ordering temperature is non-zero for
these systems, in strong contrast to the $2d$ Edwards-Anderson spin glass.

\end{abstract}
\maketitle

\mysection{Introduction}

For more than two decades the intensive numerical work on the spin glass
(SG) problem has been concentrated almost exclusively on the
Edwards-Anderson Ising spin glass : Ising spins on a regular [hyper]cubic
lattice with random near neighbor Gaussian or binomial interaction
distributions \cite{binder86}. The many possible alternative Ising systems
with a randomness ingredient have hardly been touched on and such results
as exist have been largely ignored. 

One such family of alternative systems was proposed in \cite{LC1}. It
consists of a 2d square lattice of $L\times L$ Ising spins $\sigma_i=\pm 1$
with uniform ferromagnetic
second near
 neighbor interactions of strength $J$, plus random near neighbor
interactions $J_{ij}=\pm\lambda J$; 
we will refer to it as the RCF (Randomly Coupled Ferromagnet)
model \cite{LC2}. It is described by the following Hamiltonian:
\begin{equation}
H=-\sum_{\langle i,j \rangle}J_{ij}\sigma_i \sigma_j 
-\sum_{[i,j]}J\sigma_i\sigma_j\,.
\end{equation}
For each realization of the randomness the $J_{ij}$ are drawn with the
constraint $\sum_{\langle i,j \rangle}J_{ij}=0$ to reduce
fluctuations.

As the spins
are coupled through the ferromagnetic second near neighbor
interactions, the system can be partitioned into two inter-penetrating
sublattices
in checkerboard-like fashion.
In the limit $\lambda =0$ the two sublattices
order ferromagnetically and independently below the Onsager
temperature $T=2.27 J$.  Because each sublattice can order up or down,
there are four degenerate ground states. As was pointed out in \cite{LC1},
for non-zero $\lambda$ the near neighbor interactions can be considered in
terms of effective random fields exerted by each sublattice on the other,
so that for finite $\lambda$ and large enough $L$ the ferromagnetic
sublattice ordering is expected to be broken up, as in the $2d$ random
field Ising (RFI) model \cite{RFI}. The ground state will consist of
coexisting domains of each of the four types : up/up, up/down, down/up and
down/down. The question is: is the break-up accompanied by paramagnetic
order down to $T=0$ ?

A number of Monte Carlo simulations were performed, and
it was concluded  on the basis of standard numerical criteria
\cite{LC1,LC2} that when the ratio $\lambda$ is less than about $1$,
the RCF systems show spin glass like ordering at a {\it finite} temperature,
whereas 2d Edwards Anderson ISGs are paramagnetic down to $T=0$
\cite{kawashima1997}. 
The finite ordering temperature interpretation was strongly
questioned by Parisi et al \cite{Par} who criticized the initial work
\cite{LC1} on the grounds that the results were restricted to relatively
small sample sizes L. On the basis of Monte Carlo data obtained on rather
larger samples Parisi et al suggested that the RCF systems are always
paramagnetic down to $T=0$, like the Edwards Anderson ISGs. Further large
sample Monte carlo results \cite{LC2} however indicated finite-temperature
ordering.

Here we present data from ground-state configuration evaluations
which show unambiguously that RCF systems indeed exhibit finite
temperature SG like ordering for $\lambda$ less than about $1$. This opens
up new and intriguing possibilities for the testing of fundamental
properties of complex ordered systems at finite temperatures in a $2d$
context.

\mysection{Algorithm}

In the present work,  ground state configurations have been found for
periodic boundary conditions, and for the case where in one direction
the boundary conditions are switched to anti-periodic. 
By comparing the ground state energies of the different
boundary conditions for each realization conclusions on the ordering
behavior can be obtained. Similar studies were performed for simple
$d$-dimensional EA spin glasses in $d=2$ \cite{kawashima1997}, $d=3$
\cite{alex-stiff}, and $d=4$ \cite{alex-stiff4d}.

For readers not familiar with the calculation of spin-glass ground states now
a short introduction to the subject and a description of the algorithm used
here are given. A detailed overview can be found in \cite{rieger98}

The concept of  {\em frustration} \cite{toulouse77} is important for
understanding the behavior of $\pm J$  Ising spin glasses. 
The simplest example of a frustrated system is a triple
of spins where all pairs are connected by antiferromagnetic bonds,
see fig. \ref{fig_triangle}. 
A bond is called {\em satisfied} if it contributes with a negative value
to the total energy by
choosing the values of its adjacent spins properly.
For the triangle it is
not possible to find a spin-configuration were all bonds are satisfied.
In general a system is frustrated if closed
loops of bonds exists, where the product of these bond-values is negative.
For square and cubic systems the smallest closed loops consist of four
bonds. They are called (elementary) {\em plaquettes}.

As we will see later the presence of frustration makes the 
calculation of exact ground states
of such systems computationally hard.
Only for the special case of the two-dimensional system with
periodic boundary conditions in no more than one direction and without
external field a polynomial-time algorithm is known
\cite{barahona82b}.  In general only methods with exponential running
times are known, on says the problem is {\em NP-hard}
\cite{barahona1982}. Now  for the general case three basic methods 
are briefly reviewed  and the
largest system sizes which can be treated are given 
for three-dimensional systems, the standard
spin-glass model, were data for comparison is available.

The simplest method works by
enumerating all $2^N$ possible states and has obviously an exponential
running time. Even a system size of $4^3$ is too large.  The basic idea 
of the so
called {\em Branch-and-Bound} algorithm \cite{hartwig84} is to exclude
the parts of the state space, where no low-lying states can be found, so
that the complete low-energy landscape of systems of size $4^3$ 
can be calculated \cite{klotz}. 

A more
sophisticated method called {\em Branch-and-Cut} \cite{simone95,simone96}
works by rewriting the quadratic energy function  as a linear function
with
an additional set of inequalities which must hold for the feasible solutions.
Since not all inequalities are known a priori the method iteratively
solves the linear problem, looks for inequalities which are violated,
and adds them to the set until the solution is found. Since the number
of inequalities grows exponentially with the system size the same
holds for the computation time of the algorithm. With Branch-and-Cut
anyway small systems
up to $8^3$ are feasible.

 The method used here is able to
calculate true ground states \cite{alex-stiff} up to size $14^3$. 
For two-dimensional systems, as considered in this paper,
 sizes up to $50^2$ can be
treated. The method is based on a special genetic
algorithm \cite{pal96,michal92} and on  {\em Cluster-Exact Approximation}  
\cite{alex2}. CEA is  an optimization method designed specially 
for spin glasses. Its basic idea is to transform the spin glass in a
way that graph-theoretical methods can be applied, which work only for systems
exhibiting no frustrations.
Next a description  of the genetic CEA is given.

Genetic algorithms are biologically motivated. An optimal
solution is found by treating many instances of the problem in
parallel, keeping only better instances and replacing bad ones by new
ones (survival of the fittest).
The genetic algorithm starts with an initial population of $M_i$
randomly initialized spin configurations (= {\em individuals}),
which are linearly arranged using an array. The last one is also neighbor of
the first one. Then $n_o \times M_i$ times two neighbors from the population
are taken (called {\em parents}) and two new configurations called
{\em offspring} are created. For that purpose the {\em triadic crossover}
is used which turned out to be very efficient for spin glasses: 
a mask is used which is a
third randomly chosen (usually distant) member of the population with
a fraction of $0.1$ of its spins reversed. In a first step the
offspring are created as copies of the parents. Then those spins are selected,
 where the orientations of the
first parent and the mask agree \cite{pal95}. 
The values of these spins
are swapped between the two offspring. Then a {\em mutation}
 with a rate of $p_m$
is applied to each offspring, i.e. a fraction $p_m$ of the
spins is reversed.

Next for both offspring the energy is reduced by applying
CEA:
The method constructs iteratively and randomly 
a non-frustrated cluster of spins.
Spins adjacent to many unsatisfied bonds are more likely to be added to the
cluster. During the construction of the cluster a local gauge-transformation
of the spin variables is applied so that all interactions between cluster
spins become ferromagnetic.

Fig. \ref{fig_cea_example} shows an example of how the construction 
of the cluster works using a small spin-glass system.
For 2d $\pm J$ spin glasses each cluster
contains typically 70 percent of all spins.
The  non-cluster spins act like local magnetic fields on the cluster spins,
so the ground state of the cluster is not trivial.
Since the cluster has only ferromagnetic interactions, 
an energetic minimum state for its spins can be  calculated 
in polynomial time by using graph theoretical methods 
\cite{claibo,swamy}: an equivalent network is constructed
\cite{picard1}, the maximum flow \cite{traeff,tarjan} is calculated 
\footnote{Implementation details: We used 
Tarjan's wave algorithm together
with the heuristic speed-ups of Tr\"aff. In the construction of 
the {\em level graph} we allowed not only edges $(v, w)$
with level($w$) = level($v$)+1, but also all edges $(v,t)$ where $t$
is the sink. For this measure, we observed an additional speed-up of
roughly factor 2 for the systems we calculated.} and the spins of the
cluster are set to their orientations leading to a minimum in energy. 
This minimization step
is performed $n_{\min}$ times for each offspring.

Afterwards each offspring is compared with one of its parents. The
pairs are chosen in the way that the sum of the phenotypic differences
between them is minimal. The phenotypic difference is defined here as the
number of spins where the two configurations differ. Each
parent is replaced if its energy is not lower (i.e. not better) than the 
corresponding offspring.
After this whole step is done $n_o \times M_i$ times, the population
is halved: From each pair of neighbors the configuration 
 which has the higher energy is eliminated. If more than 4
individuals remain the process is continued otherwise it
is stopped and the best individual
is taken as result of the calculation.

The representation in fig. \ref{fig_algo} summarizes the algorithm. 

The whole algorithm is performed $n_R$ times and all configurations
which exhibit the lowest energy are stored, resulting in $n_g$ statistical
independent ground state configurations. The running time of the
algorithm with suitable parameters chosen (see Table I)
grows exponentially with the
system size. On a 80Mhz PowerPC processor a typical $L=40$ instance
takes 3 hours (15 hours for $L=56$).


\mysection{Results}

In this work ground states of the RCF are studied for system sizes up
to $L=56$ and values of $\lambda=0.5$, $0.7$, $0.9$, and
$1.1$. Usually 1000 different realizations were treated, each
submitted to periodic (pbc) and antiperiodic (apbc) boundary conditions in one
direction and always pbc in the other direction. The apbc are realized
by inverting one line of bonds in the system with pbc.
Because of the enormous
computational effort, for the largest system sizes only 
realizations with $\lambda=0.7$ where considered with large statistics
(and about 100 realizations with $L=56,\lambda=0.9$).

The periodic ground states give a direct measurement of the $T=0$ break up
length $L_b$ at each value of $\lambda$, which is defined as follows:
 For small enough $L$ the ground
states will always be such that there is a full ferromagnetic ordering
within each sublattice. With increasing $L$, more and more samples will be
found with ground states having at least one of the sublattices
incompletely ferromagnetic. The break up length $L_b$ is defined \cite{Sep}
as the value of $L$ above which more than half the samples do not have pure
ferromagnetic order in each sublattice.  For the binomial RFI model, $L_b
\sim 5.5\exp (2/\Delta^2)$ where $\Delta $ is the strength of the random field
\cite{Sep}. For the RCF the values of $L_b$ are shown against
$\lambda^{-2}$ in Figure 1. It was suggested in \cite{LC2} that by analogy
with the RFI results \cite{Sep} $L_b(\lambda)$ could be expected to vary as
$\exp (1/(4\lambda^2))$. In fact the data points for the true ground states
lie on the line $L_b \sim3.2\exp (0.62/\lambda^2)$. For the particular cases
$\lambda=0.5$ and $\lambda=0.7$, $L_b\approx 45$ and $\approx 10$ 
respectively. With the
wisdom of hindsight, it can be seen that the measurements done in
\cite{LC1,LC2,Par} for $\lambda=0.5$ were mainly in the regime $L <L_b$
while for $\lambda= 0.7$ the larger samples were well in the regime $L >
L_b$. 

A ``typical'' ground state for $\lambda = 0.7$ and $L=56$ is shown in
Figure \ref{figSampleState}. All four possible types of domains occur.
Because of the discrete structure of the interaction usually the
ground state is degenerate. But in contrast to the EA spin glasses
with only $\pm J$ near neighbor 
interactions, where a complex ground-state landscape exists,
the structure of the degeneracy is
trivial for $\lambda \le 1$:
the whole system may be flipped, sometimes it is possible to
flip both sublattices independently, and usually some small clusters occur
with can take two orientations.

But for studying whether the model exhibits long range order or not,
 it is sufficient to concentrate on the
ground-state energies $E_P,E_{AP}$ 
for periodic and antiperiodic boundary conditions.
The energy differences $\Delta  = E_P-E_{AP}$ give information about
whether a system exhibits some kind of stiffness against perturbations
of the boundary, i.e. about the presence of order. $\Delta$ is called
the {\em stiffness energy}.
For samples with the same set of interactions
the stiffness can be analyzed in terms of the size dependence of the average
  $\langle\Delta \rangle$ 
and of the width $W\equiv \sqrt{\sigma^2(\Delta)}$ of the
distribution $P(\Delta)$. For $\lambda=0.7$ the distribution 
is presented in Fig. \ref{figDistrDeltaMean}. The inset shows
the behavior of the average stiffness energy as a function of $L$ for all four
values of $\lambda$. For system sizes larger the breakup length
and $\lambda \ge 0.7$ the
stiffness energy decreases, indicating that no ferromagnetic long
range order is present in the system. For $\lambda=0.5$ the breakup
length is very large, so the asymptotic behavior is hardly visible,
but $\langle \Delta \rangle$ seems to fall for $L\ge 28$.
From direct evaluation of the magnetization (see Fig
\ref{figMagLambda} and Fig. \ref{figLambdaKL}) we conclude
that no ferromagnetic order should be present beyond an upper limit
$\lambda=0.27(8)$. For smaller values of $\lambda$ nothing can be
concluded from our data. Furthermore, for smaller values of $\lambda$
it remains possible that the ground  
states of the RCF
model do not exhibit ferromagnetic ordering
 ordering for any finite value of the
relative coupling constant $\lambda$.

From standard relationships \cite{bray1984,mcmillan1984} one can write 
$\langle\Delta \rangle \sim L^{\theta_F}$ with $\theta_F$ the ferromagnetic
stiffness exponent, and $W \sim L^{\theta_{SG}}$ with $\theta_{SG}$ the
spin glass stiffness exponent.  Positive values of the exponents
indicate a long range order.
Because of the small system sizes an evaluation of the ferromagnetic
exponent is difficult. From the results presented in Fig. 
\ref{figDistrDeltaMean} we find
an asymptotic ($L\to\infty$) value of $\theta_F=-2$ ($\lambda=0.9, 1.1$). 

Now we turn to the question whether some kind of spin-glass order is
present in the system. This can be investigated by analyzing
The dependence of the variance $\sigma^2(\Delta)$ of 
the stiffness-energy distributions 
on the system size, the result is
shown in Fig. \ref{figWidthDelta}. For small system sizes the variance
grows for all values of the coupling constant $\lambda$. In order to
exclude finite-size effects, only systems larger than the breakup
length $L_b(\lambda)$ should be taken into account.
Above $L_b$ there is
a good linear size dependence of $\log W(L)$ against $\log L$, with
$\theta_{SG}=0.59(8)$, $0.29(1)$, $0.09(5) $, and $-0.16(2)$ 
respectively for   $\lambda = 0.5, 0.7,0.9$, and $1.1$. 

The values of $ \theta_{SG}$ against $\lambda$ are shown in the inset
of Figure \ref{figWidthDelta}. The result for
$\lambda=0.5$ is not very reliable, because the largest system size is of the
order of the breakup length. In the log-log plot the datapoints for
$\lambda=0.5$ 
exhibit a negative curvature, thus the asymptotic value of
$\theta_{SG}$ may be smaller than $0.59$.
For the other systems the
breakup length is quite small, so the results give
 unambiguous evidence for spin glass like ordering in the large
size limit, with a non-zero ordering temperature. 
Especially for $\lambda=0.7$, where $L_b\approx 10$, the result
$\sigma(\Delta)>0$ is very reliable.
 Thus, it is indeed not necessary
to carry out further calculations with larger systems 
to prove the fact, there there are values of the
coupling constant giving rise to an ordered spin glass phase in the RCF.
The limiting value $\lambda_c$ above which $\theta_{SG}$ is negative is very
close to $1.0$; $\lambda_c$ would correspond to the  highest value at which
the ordering temperature is non-zero, in good agreement with the initial
estimate  from the Monte Carlo work \cite{LC1}.

\mysection{Conclusion}

We have calculated ground states of the Randomly Coupled Ferromagnet
for different values of the spin-glass coupling constant $\lambda$ and
with periodic as well as antiperiodic boundary conditions. By using
the genetic cluster-exact approximation algorithm, we were able to
treat system sizes up to $N=56\times 56$. The breakup length was
calculated for each value of $\lambda$. From the calculation of 
$T=0$ stiffness energy it could be concluded that below $\lambda_c\approx 1$
the RCF exhibits an ordered spin glass like phase at finite
temperature. It should be stressed again that for $\lambda>0.5$ the
largest system sizes are well beyond the breakup length, so no
changes are to be expected for larger system sizes. For $\lambda<0.5$,
especially if one likes to test whether the model exhibits ferromagnetic
ordering, 
ground states calculations of larger systems are needed to study the
behavior in more detail. Unfortunatley, these studies
 are beyond the power of current computers and algorithms.

Although the zero-temperature stiffness exponent values give no direct
information on the ordering temperatures, the present results are
consistent with the the conclusions drawn in \cite{LC1,LC2} where Monte
Carlo estimates of the critical temperatures were made using the 
finite-size scaling of the spin glass susceptibility and the form of the time
dependence of the autocorrelation function relaxation. Ordering
temperatures were estimated to be close to $2.0$ for $\lambda= 0.5$ and
$0.7$, dropping to zero near $\lambda = 1$. Rather remarkably the $T=0$
crossover as a function of $L$ at $L_b$ appears to have little effect on
the behavior of the SG susceptibility as a function of size in the
temperature region close to $T_g$ \cite{LC2}. However for $\lambda = 0.5$
Parisi et al \cite{Par} observed weakly non-monotonic behavior of the
Binder parameter with $L$ for sizes that we now know to be in the
region of the crossover.

Since the existence of a spin glass like phase at finite temperature
now has been established definitely,
it would be instructive to carry out
further careful Monte Carlo measurements for sample sizes well in the
regime $L > L_b$ and over a range of $\lambda$ values. Is the physics of
the $2d$ RCF above, at, and below the ordering temperature strictly
analogous to that of the standard Edwards Anderson ISG at dimensions where
there is finite temperature ordering? To what extent could the RCF
enlighten us concerning problems which in the Edwards Anderson ISG context
have remained conflictual for more than twenty years ? The fact that the
RCF lives on a $2d$ lattice rather than in a higher dimension should
facilitate understanding of the fundamental physics of ordering in complex
systems.

Finally, there may even be possible experimental realizations of systems 
where quasi-two-dimensional magnets form short range clusters with local
ferromagnetic
or antiferromagnetic order, with random frustrated interactions linking
these clusters 
together. Examples of promising behaviour of this sort are Fe compound with
halogens
\cite{fecl} where it might be interesting to look at the data again 
in view of the present results.

\mysection{Acknowledgements}

AKH was supported by the Graduiertenkolleg
``Modellierung und Wissenschaftliches Rechnen in 
Mathematik und Naturwissenschaften'' at the
{\em In\-ter\-diszi\-pli\-n\"a\-res Zentrum f\"ur Wissenschaftliches
  Rechnen} in Heidelberg and the
{\em Paderborn Center for Parallel Computing}
 by the allocation of computer time.  AKH obtained financial
 support from the DFG ({\em Deutsche Forschungs Gemeinschaft}) under
 grant Zi209/6-1.
IAC gratefully acknowledges very helpful discussions with Dr N. Lemke,
and
thanks
Professor T. Shirakura for having shown him very interesting unpublished data.
Meetings organized by the Monbusho collaboration "Statistical physics of
fluctuations in glassy systems" and by the ESF network "Sphinx" played an
essential role for the present work.

\begin{table}[ht]
\begin{center}
\begin{tabular}{cccccc}
\hline
$L$ & $M_i$ & $n_{o}$ & $n_{\min}$ & $p_m$  & $n_R$ \\ \hline
5 & 8 & 1 & 1 & 0.05 & 5\\
10 & 16 & 1 & 2 & 0.05 & 5\\
14 & 16 & 2 & 2 & 0.05 & 5 \\
20 & 32 & 8 & 2 & 0.05 & 5\\
28 & 128 & 16 & 2 & 0.05 & 5\\
40 & 512 & 16 & 2 & 0.05 & 5\\
56 & 1024 & 16 & 2 & 0.05  & 5
\end{tabular}
\end{center}
\caption{Simulation parameters: $L$ = system size, $M_i$ = initial size of
population, $\nu$ = average number of offsprings per configuration, $n_{\min}$
= number of CEA minimization steps per offspring, $p_m$ = mutation
rate, $n_R$=number of independent runs per realization.}
\label{tab_parameters}
\end{table}

\newcommand{\captionK}
{The simplest frustrated system: a triple of spins, each pair of spins
connected by antiferromagnetic bonds (dashed lines). It is not
possible to satisfy all bonds.
}

\newcommand{\captionALGO}
{Genetic Cluster-exact Approximation.
}

\newcommand{\captionA}
{Example of the Cluster-Exact Approximation method. A part of a spin glass
is shown. The circles represent lattice sites/spins. Straight lines represent
ferromagnetic bonds the jagged lines antiferromagnetic interactions. The
top part shows the initial situation. 
The construction starts with the spin at the center. The bottom part 
displays the final stage.
The spins which belong to the cluster carry a plus or minus sign which
indicates how each spin is transformed, so that only ferromagnetic
interactions remain inside the cluster. All other spins cannot be added
to the cluster because it is not possible to multiply them by $\pm 1$
to make all adjacent bonds positive. Please note that many other combinations
of spins can be used to build a cluster without frustration.}

\newcommand{\captionLBLambda}
{Breakup length $L_b$ as a function of $1/\lambda^2$. The solid line shows a
  fit to the function $L_b(\lambda)=a\exp(-b/\lambda^2)$.}


\newcommand{\captionMagLambda}
{Average magnetization $m$ as a function of the strength $\lambda$ of
  the spin-glass couplings for different systems sizes $5\le L \le 56$.}

\newcommand{\captionLambdaKL}
{Threshold value $\lambda_c$ as a function of the system size $L$. The
  threshold
value is determined from $m(\lambda_c)=0.9$. The solid line shows a
fit to a function $\lambda_c(L)=\lambda_{\infty}+eL^{-f}$, resulting
in $\lambda_{\infty}=0.27(8)$, $f=-0.53(10)$.}

\newcommand{\captionSampleState}
{Typical ground state of one RCF realization ($L=56$) for $\lambda=0.70$ with
  periodic boundary conditions in all directions. Two different
  symbols (white/black square), (unfilled/filled diamond) are used to
  represent the orientations on the different sublattices.}

\newcommand{\captionDistrDeltaMean}
{Distribution of stiffness energies $\Delta$ at $\lambda=0.7$ for
  different system sizes $L=14,28,56$. The inset shows the mean
  stiffness energy $\langle \Delta \rangle$ 
as a function of $L$ for $\lambda=0.5$,  $0.7$,  $0.9$, and $1.1$.}

\newcommand{\captionWidthDelta}
{Width $\sigma^2(\Delta)$ of the distribution of stiffness energies as
  a function of the system size $L\ge 10$ for $\lambda=0.5, 0.7, 0.9,
  1.1$. The solid
  lines are fits to algebraic functions of the form
  $\sigma^2(L)=gL^{\theta_{SG}}$. The inset shows
 the values of the exponent for different values of $\lambda$.}



\newpage

\begin{figure}[htb]
\begin{center}
\myscaleboxb{\includegraphics{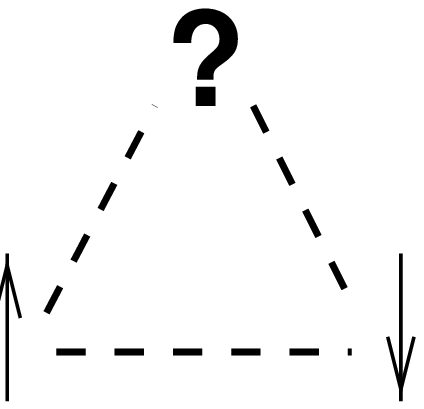}}
\end{center}
\caption{\captionK}
\label{fig_triangle}
\end{figure}

\begin{figure}[htb]
\begin{center}
\myscaleboxc{\includegraphics{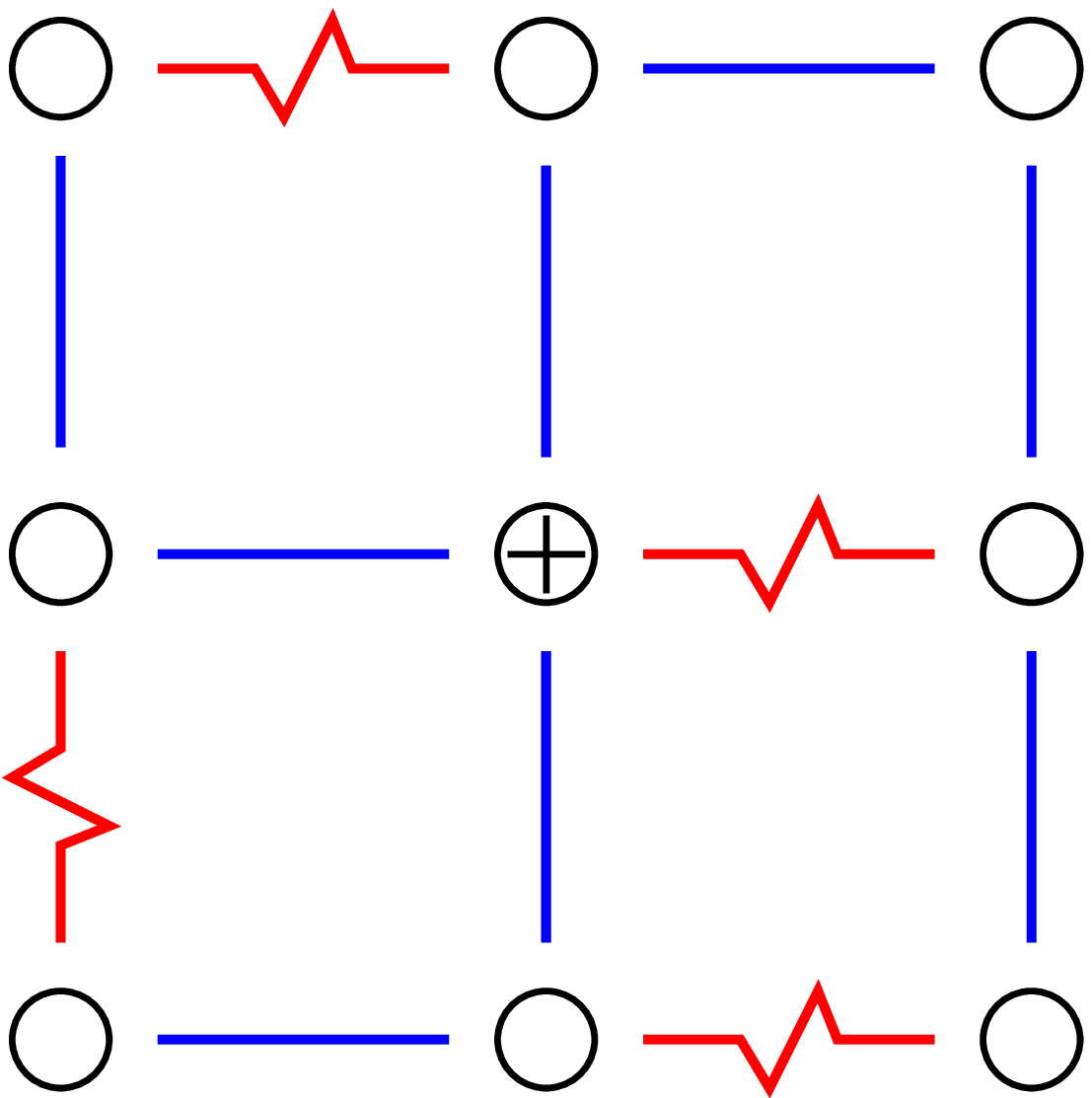}}
\vspace{0.2cm}
\myscaleboxc{\includegraphics{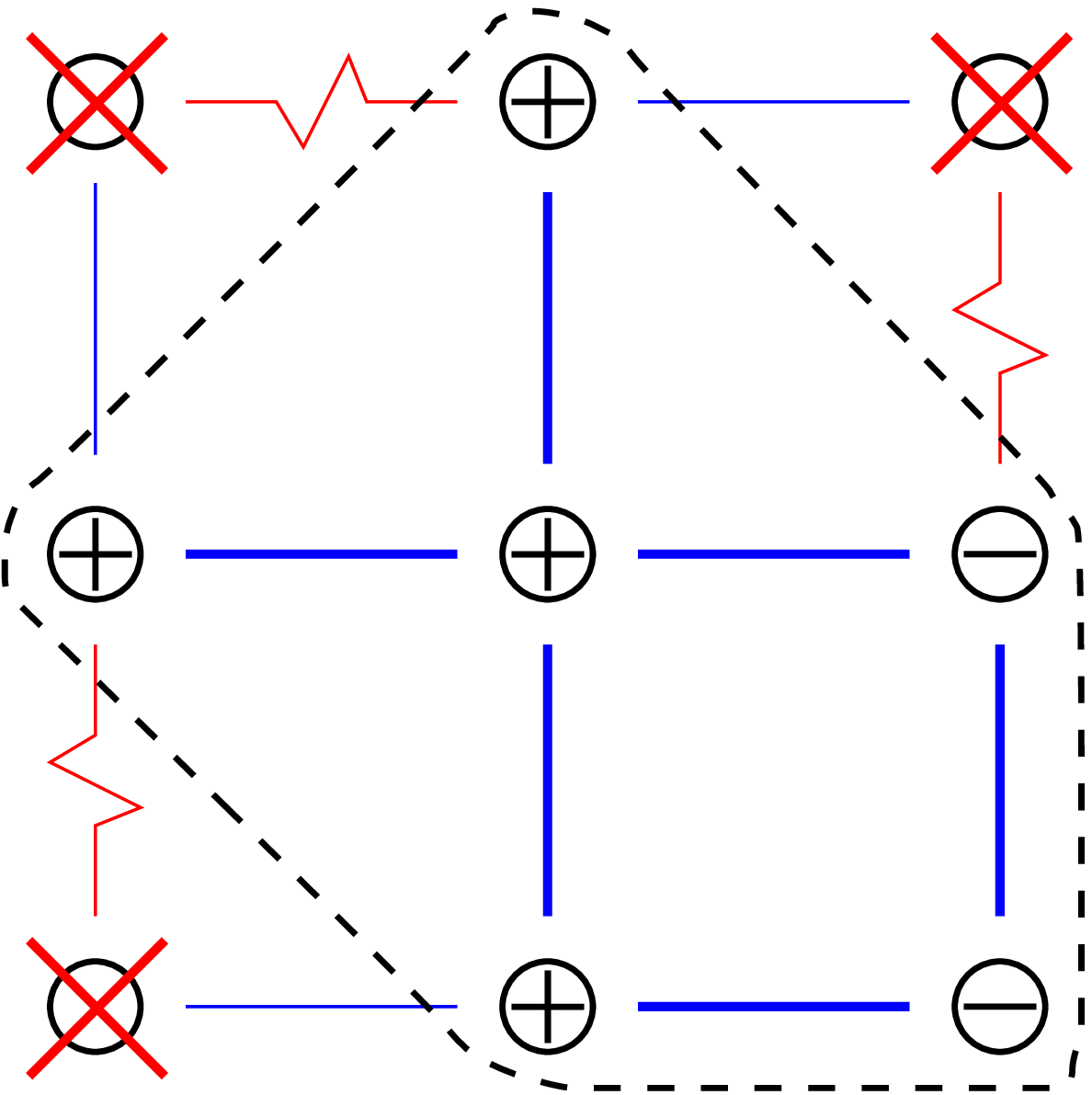}}
\end{center}
\caption{\captionA}
\label{fig_cea_example}
\end{figure}

\newlength{\mpwidth}
\setlength{\mpwidth}{\textwidth}
\addtolength{\mpwidth}{-2cm}
\begin{figure}
\begin{center}
\begin{minipage}[b]{\mpwidth}
\newlength{\tablen}
\settowidth{\tablen}{xxx}
\newcommand{\tabspace}{\hspace*{\tablen}}
\begin{tabbing}
\tabspace \= \tabspace \= \tabspace \= \tabspace \= \tabspace \=
\tabspace \= \kill
{\bf algorithm} genetic CEA($\{J_{ij}\}$,
$M_i$, $n_o$, $p_m$, $n_{\min}$)\\
{\bf begin}\\
\> create $M_i$ configurations randomly\\
\> {\bf while} ($M_i > 4$) {\bf do}\\
\> {\bf begin}\\
\> \> {\bf for} $i=1$ {\bf to} $n_o \times M_i$ {\bf do}\\
\>\> {\bf begin}\\
\>\>\> select two neighbors \\
\>\>\> create two offspring using triadic crossover\\
\>\>\> do mutations with rate $p_m$\\
\>\>\> {\bf for} both offspring {\bf do}\\
\>\>\> {\bf begin}\\
\>\>\>\> {\bf for} $j=1$ {\bf to} $n_{\min}$ {\bf do}\\
\>\>\>\> {\bf begin}\\
\>\>\>\>\> construct unfrustrated cluster of spins\\
\>\>\>\>\> construct equivalent network\\
\>\>\>\>\> calculate maximum flow\\
\>\>\>\>\> construct minimum cut\\
\>\>\>\>\> set new orientations of cluster spins\\
\>\>\>\> {\bf end}\\
\>\>\>\> {\bf if} offspring is not worse than related parent \\
\>\>\>\> {\bf then}\\
\>\>\>\>\> replace parent with offspring\\
\>\>\> {\bf end}\\
\>\> {\bf end}\\
\>\> half population; $M_i=M_i/2$\\
\> {\bf end}\\
\> {\bf return} one configuration with lowest energy\\
{\bf end}
\end{tabbing}
\end{minipage}
\end{center}
\caption{\captionALGO}
\label{fig_algo}
\vspace{0.5cm}
\end{figure}

\begin{figure}[htb]
\begin{center}
\myscalebox{\includegraphics{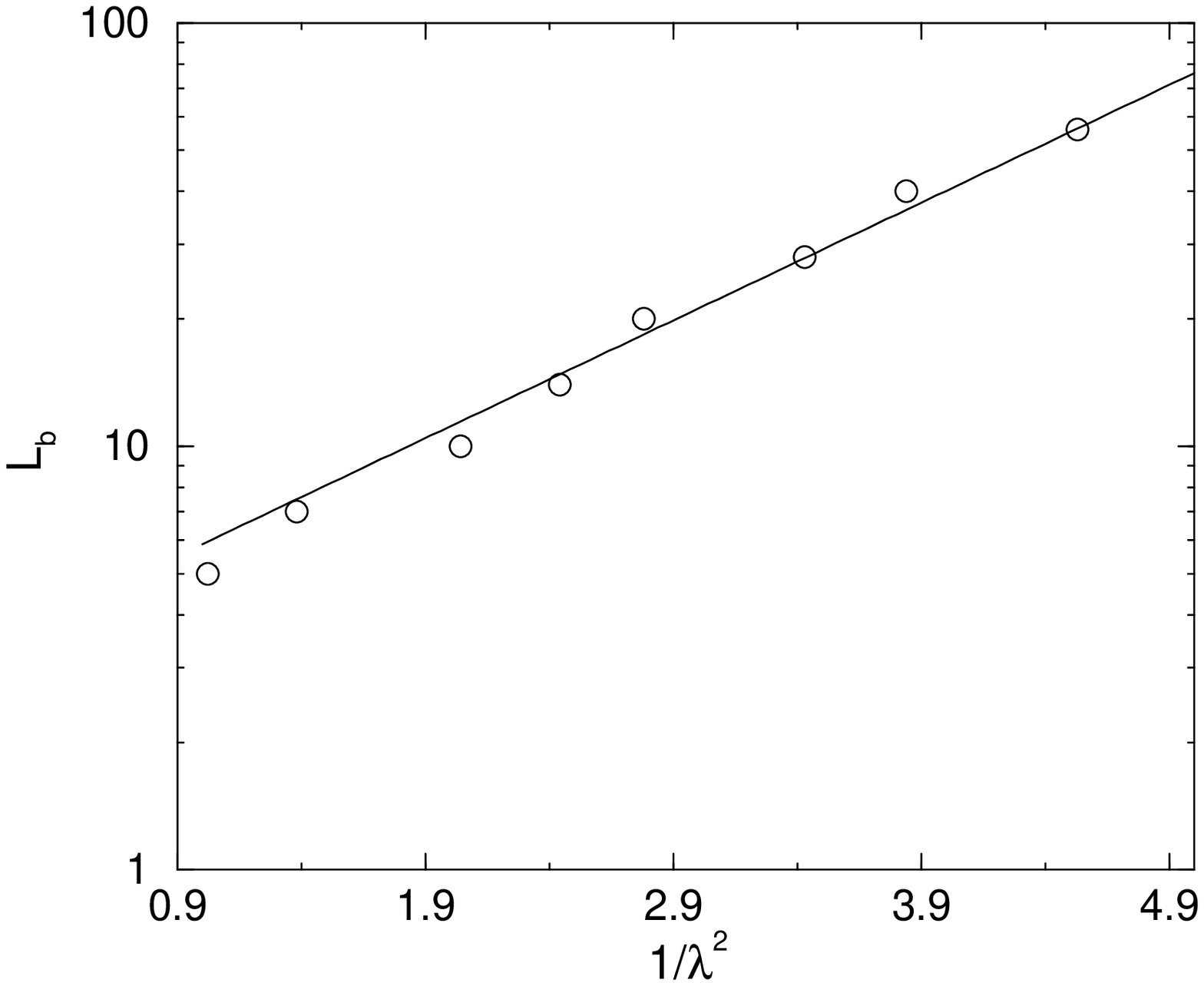}}
\end{center}
\caption{\captionLBLambda}
\label{figLBLambda}
\end{figure}

\begin{figure}[htb]
\begin{center}
\myscaleboxb{\includegraphics{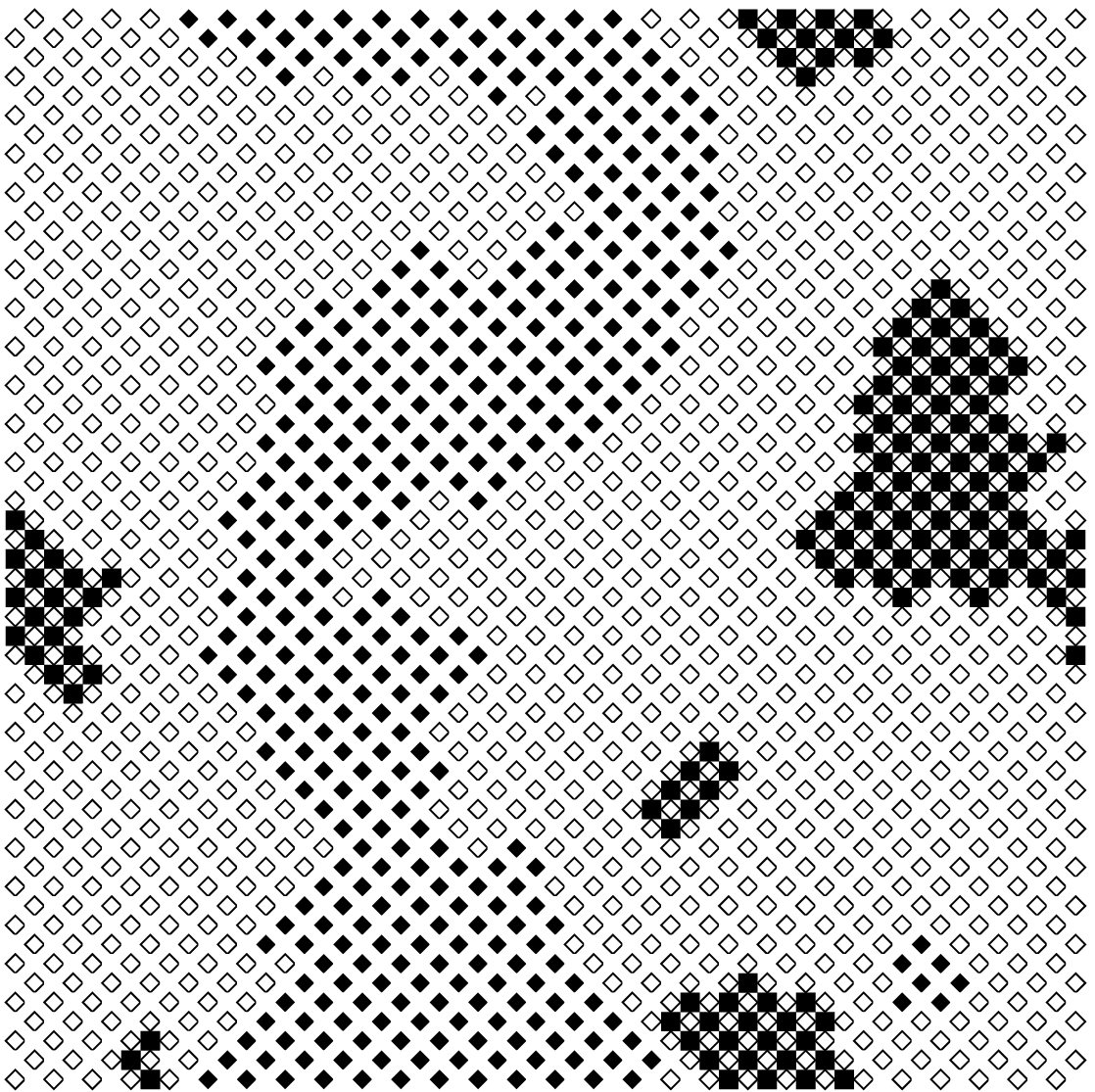}}
\end{center}
\caption{\captionSampleState}
\label{figSampleState}
\end{figure}

\begin{figure}[htb]
\begin{center}
\myscalebox{\includegraphics{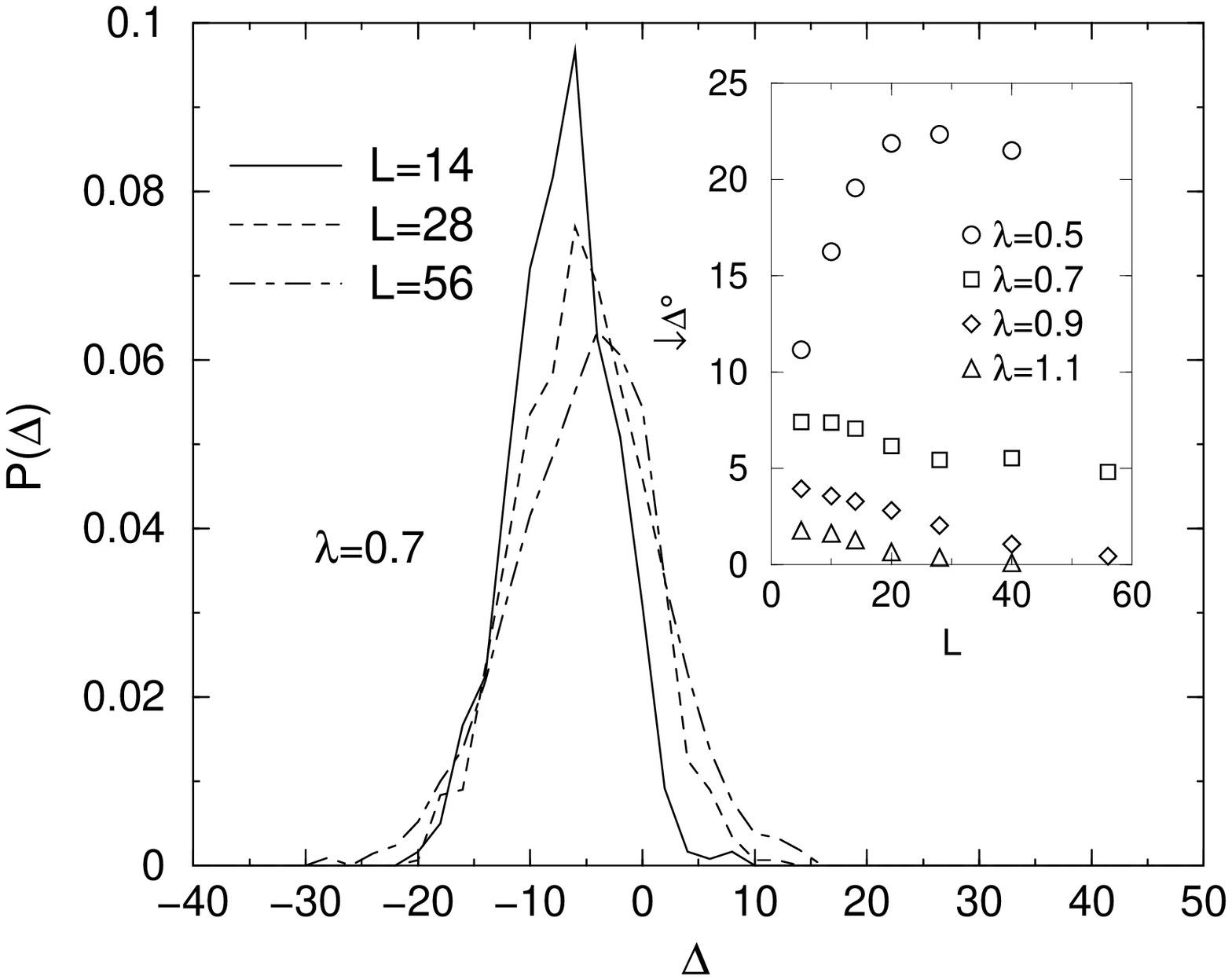}}
\end{center}
\caption{\captionDistrDeltaMean}
\label{figDistrDeltaMean}
\end{figure}

\begin{figure}[htb]
\begin{center}
\myscalebox{\includegraphics{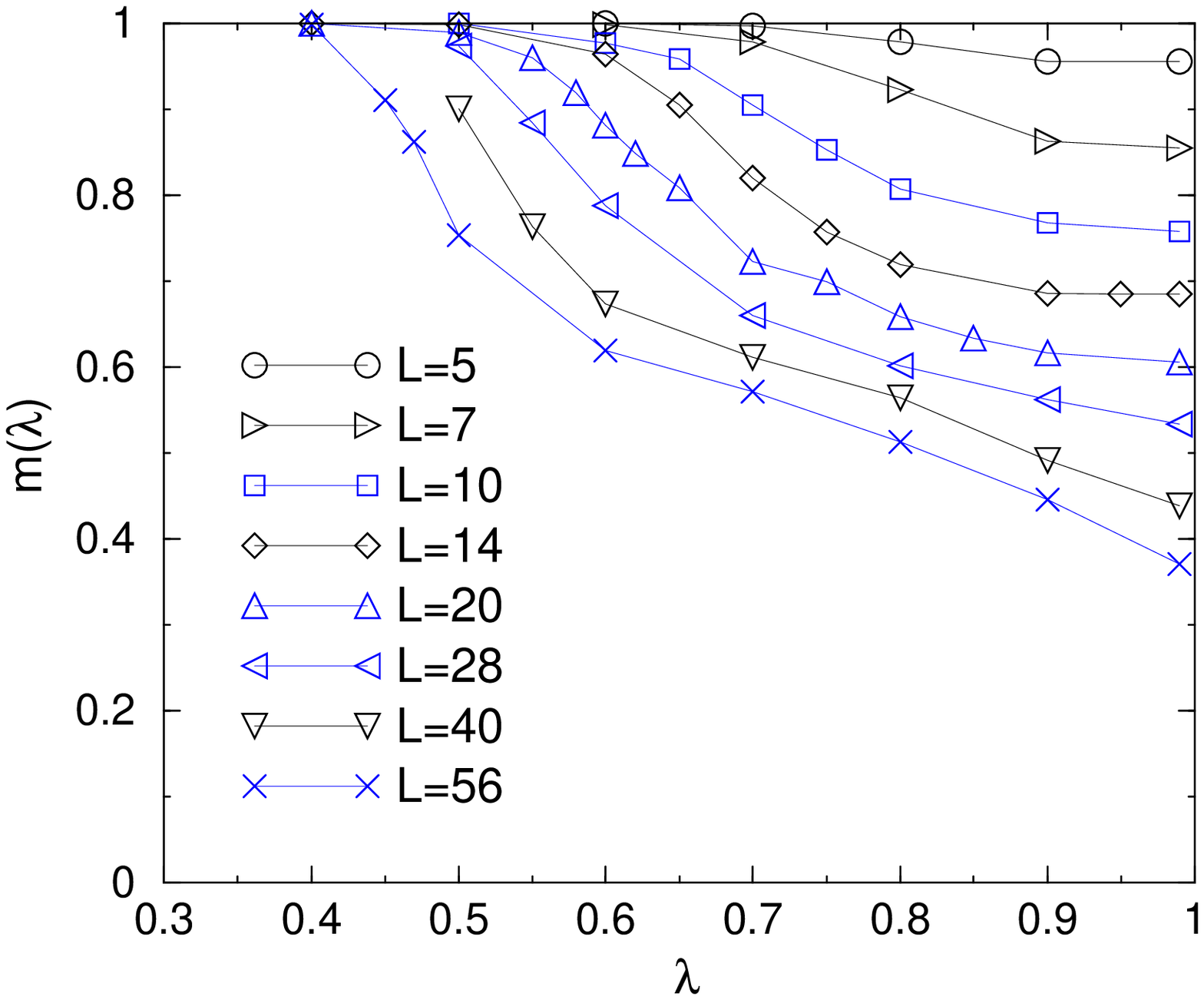}}
\end{center}
\caption{\captionMagLambda}
\label{figMagLambda}
\end{figure}

\begin{figure}[htb]
\begin{center}
\myscalebox{\includegraphics{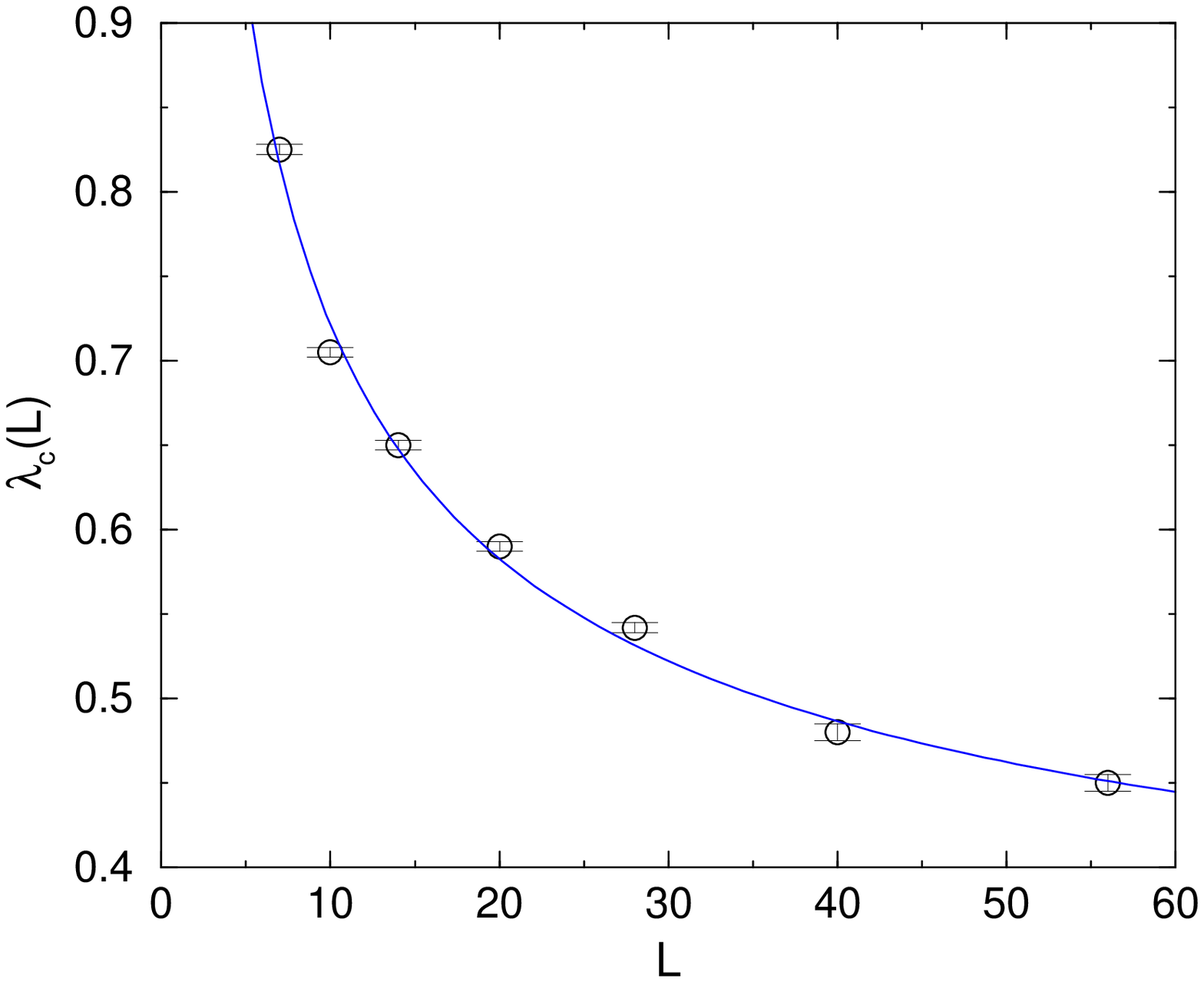}}
\end{center}
\caption{\captionLambdaKL}
\label{figLambdaKL}
\end{figure}

\begin{figure}[htb]
\begin{center}
\myscalebox{\includegraphics{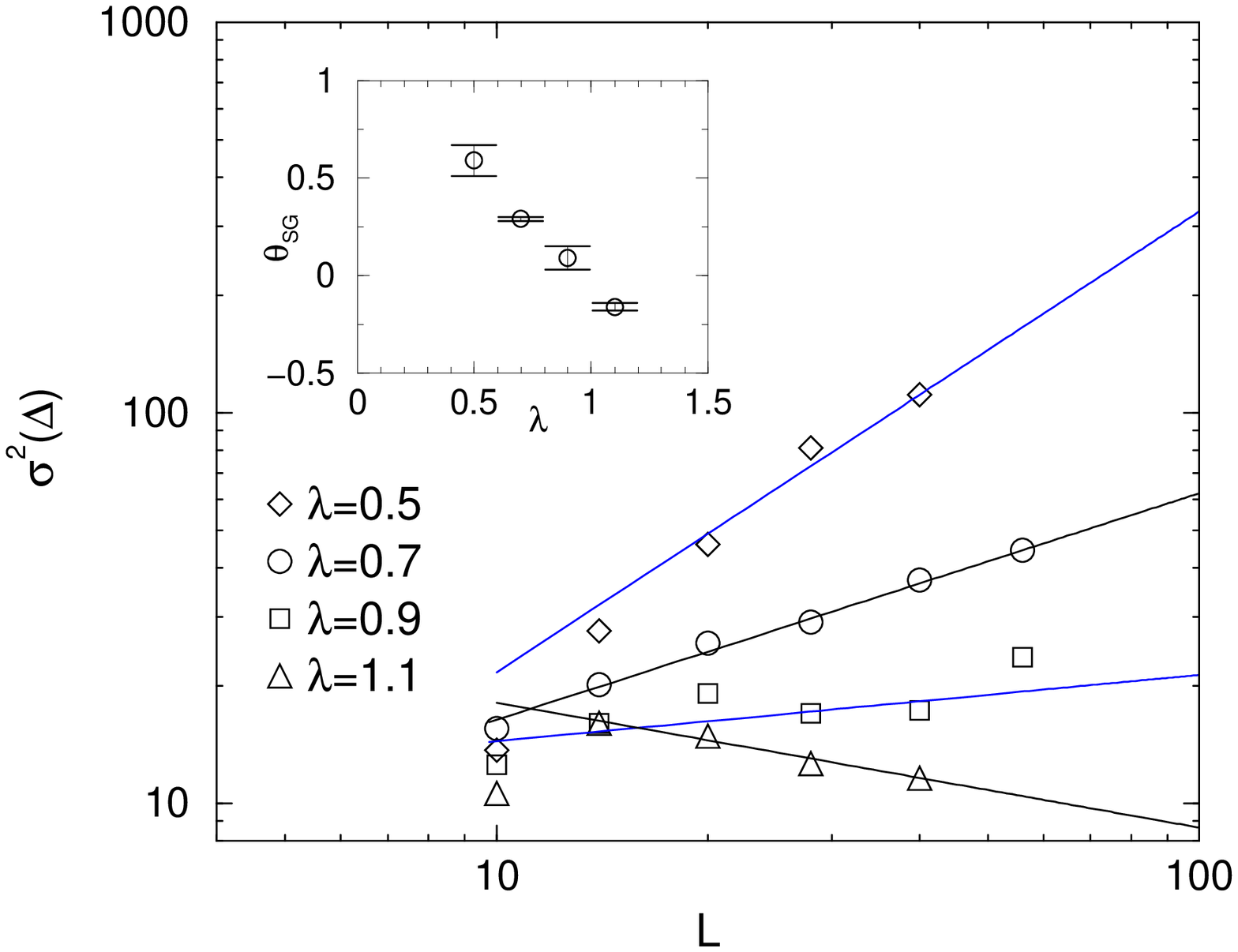}}
\end{center}
\caption{\captionWidthDelta}
\label{figWidthDelta}
\end{figure}


\begin{thebibliography}{99}

\bibitem{binder86} For reviews on spin glasses see: 
K. Binder and A.P. Young, Rev. Mod. Phys. {\bf 58}, 801 (1986);
 M. Mezard, G. Parisi, M.A. Virasoro, Spin
glass theory and beyond, World Scientific, Singapore 1987;
 K.H. Fisher and J.A. Hertz, Spin Glasses, Cambridge
University Press, 1991

\bibitem{LC1}N. Lemke and I.A. Campbell, Phys. Rev. Lett. {\bf 76}, 4616 (1996)

\bibitem{LC2}N. Lemke and I.A. Campbell, J. Phys.A {\bf 32}, 7851 (1999)

\bibitem{kawashima1997} N. Kawashima and H. Rieger, 
  Europhys. Lett. {\bf 39}, 85 (1997)

\bibitem{Par}G. Parisi, J.J. Luiz-lorenzo, and D.A. Stariolo, J. Phys A {\bf
31}  4657, (1998)

\bibitem{RFI}Y. Imry and S.-K. Ma, Phys. Rev. Lett. {\bf 35}, 1399 (1975),
M. Aizenman and J. Wehr, Phys. Rev. Lett. {\bf 62}, 2503 (1989)

\bibitem{alex-stiff} A.K. Hartmann, Phys. Rev. E {\bf 59}, 84 (1999)

\bibitem{alex-stiff4d} A.K. Hartmann, Phys. Rev. E {\bf 60}, 5135 (1999) 

\bibitem{rieger98} H. Rieger, in: Advances in Computer Simulation, 
     ed. J. Kertesz and I. Kondor,
    Lecture Notes in Physics {\bf 501}, (Springer-Verlag, Heidelberg, 1998)

\bibitem{toulouse77} G. Toulouse, Commun. Phys. {\bf 2}, 115 (1977)

\bibitem{barahona82b} F. Barahona, R. Maynard, R. Rammal and
  J.P. Uhry, J. Phys. A {\bf 15}, 673 (1982).

\bibitem{barahona1982} F. Barahona, J. Phys. A {\bf 15}, 3241 (1982)

\bibitem{hartwig84} A. Hartwig, F. Daske and S. Kobe, 
Comp. Phys. Commun. {\bf 32} 133 (1984)

\bibitem{klotz} T. Klotz and S. Kobe, J. Phys. A: Math. Gen. {\bf 27},
L95 (1994)

\bibitem{simone95} C. De Simone, M. Diehl, M. J\"unger, P. Mutzel, 
G. Reinelt and G. Rinaldi, J. Stat. Phys. {\bf 80}, 487 (1995)

\bibitem{simone96} C. De Simone, M. Diehl, M. J\"unger, P. Mutzel, 
G. Reinelt and G. Rinaldi, J. Stat. Phys. {\bf 84}, 1363 (1996)

\bibitem{pal96} K.F. P\'al, Physica A {\bf 223}, 283 (1996)

\bibitem{michal92} Z. Michalewicz, Genetic Algorithms + Data Structures
= Evolution Programs, Springer, Berlin 1992

\bibitem{alex2} A.K. Hartmann, Physica A {\bf 224}, 480 (1996)

\bibitem{pal95} K.F. P\'al, Biol. Cybern. {\bf 73}, 335 (1995)

\bibitem{claibo}J.D. Claiborne, Mathematical Preliminaries for
Computer Networking, Wiley, New York 1990

\bibitem{swamy} M.N.S. Swamy and K. Thulasiraman, Graphs, Networks and
Algorithms, Wiley, New York 1991

\bibitem{picard1} J.-C. Picard and H.D. Ratliff, Networks {\bf 5}, 357 (1975)

\bibitem{traeff} J.L. Tr\"aff, Eur.\ J.\ Oper.\ Res.\ {\bf 89}, 564 (1996)

\bibitem{tarjan} R.E. Tarjan, Data Structures and Network Algorithms,
Society for industrial and applied mathematics, Philadelphia 1983




\bibitem{Sep}E.T. Sepp\"al\"a, V. Pet\"aj\"a and M.J. Alava, Phys.Rev.E {\bf
58}  5217, (1998)

\bibitem{bray1984} A.J. Bray and M.A. Moore, J. Phys. C {\bf 17}, L463 (1984)

\bibitem{mcmillan1984} W.L. McMillan, Phys. Rev. B {\bf 30}, 476 (1984)

\bibitem{fecl}J. Vetel, M. Yahiaoui, D. Bertrand, A.R. Fert, J.P. Redoules
and J. Ferre
J. Physique, Colloque {\bf C8}, {\bf 49} 1067 (1988),
D. Bertrand, F. Bensamka, A.R. Fert, J. Gelard, J.P. Redoules and S. Legrand
J. Phys. C: Solid State Phys. {\bf 17} 1725 (1984)


\end{thebibliography}
\end{document}